\documentclass[10pt,]{article}
\pdfoutput=1
\usepackage{booktabs}
\usepackage{lmodern}
\usepackage{amssymb,amsmath}
\usepackage{ifxetex,ifluatex}
\usepackage{fixltx2e} % provides \textsubscript
\ifnum 0\ifxetex 1\fi\ifluatex 1\fi=0 % if pdftex
  \usepackage[T1]{fontenc}
  \usepackage[utf8]{inputenc}
\else % if luatex or xelatex
  \ifxetex
    \usepackage{mathspec}
  \else
    \usepackage{fontspec}
  \fi
  \defaultfontfeatures{Ligatures=TeX,Scale=MatchLowercase}
\fi
\IfFileExists{upquote.sty}{\usepackage{upquote}}{}
\IfFileExists{microtype.sty}{%
\usepackage{microtype}
\UseMicrotypeSet[protrusion]{basicmath} % disable protrusion for tt fonts
}{}
\usepackage{hyperref}
\hypersetup{unicode=true,
            pdftitle={Efficient estimation of the maximum metabolic productivity of batch systems},
            pdfborder={0 0 0},
            breaklinks=true}
\usepackage{natbib}
\usepackage{graphicx,grffile}
\makeatletter
\def\maxwidth{\ifdim\Gin@nat@width>\linewidth\linewidth\else\Gin@nat@width\fi}
\def\maxheight{\ifdim\Gin@nat@height>\textheight\textheight\else\Gin@nat@height\fi}
\makeatother
\setkeys{Gin}{width=\maxwidth,height=\maxheight,keepaspectratio}
\IfFileExists{parskip.sty}{%
\usepackage{parskip}
}{% else
\setlength{\parindent}{0pt}
\setlength{\parskip}{6pt plus 2pt minus 1pt}
}
\usepackage[top=1in, bottom=1in, left=1in, right=1in]{geometry}

\usepackage[T1]{fontenc}
\usepackage[sc]{mathpazo}
\usepackage{textcomp}
\usepackage{inconsolata}

\usepackage[labelfont={bf}, margin=1cm]{caption}
\usepackage{titlesec}
\titleformat*{\subsubsection}{\itshape\bfseries}
\titleformat*{\paragraph}{\itshape}

\usepackage{cite}

\title{Efficient estimation of the maximum metabolic productivity of batch
systems}
\date{}

\begin{document}
\maketitle

\emph{Peter C. St.~John\textsuperscript{1}, Michael F.
Crowley\textsuperscript{1}, Yannick J. Bomble\textsuperscript{1}}

\textsuperscript{1}Biosciences Center, National Renewable Energy
Laboratory, Golden CO 80401

\section{Abstract}\label{abstract}

\subsection{Background}\label{background}

Production of chemicals from engineered organisms in a batch culture
involves an inherent trade-off between productivity, yield, and titer.
Existing strategies for strain design typically focus on designing
mutations that achieve the highest yield possible while maintaining
growth viability. While these methods are computationally tractable, an
optimum productivity could be achieved by a dynamic strategy in which
the intracellular division of resources is permitted to change with
time. New methods for the design and implementation of dynamic microbial
processes, both computational and experimental, have therefore been
explored to maximize productivity. However, solving for the optimal
metabolic behavior under the assumption that all fluxes in the cell are
free to vary is a challenging numerical task. Previous studies have
therefore typically focused on simpler strategies that are more feasible
to implement in practice, such as the time-dependent control of a single
flux or control variable.

\subsection{Results}\label{results}

This work presents an efficient method for the calculation of a maximum
theoretical productivity of a batch culture system using a dynamic
optimization framework. The proposed method follows traditional
assumptions of dynamic flux balance analysis: first, that internal
metabolite fluxes are governed by a pseudo-steady state, and secondly
that external metabolite fluxes are dynamically bounded. The
optimization is achieved via collocation on finite elements, and
accounts explicitly for an arbitrary number of flux changes. The method
can be further extended to calculate the complete Pareto surface of
productivity as a function of yield. We apply this method to succinate
production in two engineered microbial hosts, \emph{Escherichia coli}
and \emph{Actinobacillus succinogenes}, and demonstrate that maximum
productivities can be more than doubled under dynamic control regimes.

\subsection{Conclusions}\label{conclusions}

The maximum theoretical yield is a measure that is well established in
the metabolic engineering literature and whose use helps guide strain
and pathway selection. We present a robust, efficient method to
calculate the maximum theoretical productivity: a metric that will
similarly help guide and evaluate the development of dynamic microbial
bioconversions. Our results demonstrate that nearly optimal yields and
productivities can be achieved with only two discrete flux stages,
indicating that near-theoretical productivities might be achievable in
practice.

\section{Keywords}\label{keywords}

flux balance analysis, dynamic optimization, elementary flux modes,
\emph{Actinobacillus succinogenes}, \emph{Escherichia coli}

\section{Background}\label{background-1}

Microbial bioconversion plays a critical role in efforts to enable
sustainable production of commodity chemicals from renewable feedstocks.
The economic feasibility of the integrated biorefinery concept therefore
hinges sharply on the productivity, yield, and titers that can be
achieved by a given microbial host \citep{Davis:2013jl}. Flux balance
modeling has emerged as an important tool in guiding experimental
efforts in strain design by predicting the effects of gene knockouts and
overexpression on metabolite yields \citep{Feist:2009ik}. In developing
engineered organisms for optimal performance in batch cultures, A
trade-off is often encountered between the productivity and yield that
can be obtained via metabolic interventions \citep{Holtz:2010bm}. As a
result, many existing strategies for strain design involve designing
mutations to achieve the highest yield possible while maintaining growth
viability \citep{Burgard2003, Ranganathan:2010do}. Other strategies have
specifically addressed the importance of productivity through dynamic
simulations \citep{Zhuang:2013fu, Hanly:2013km}. Overall, approaches
tend to follow the principle of designing static networks with minimum
metabolic functionality to achieve desired yields
\citep{Ruckerbauer2014}. While these methods are computationally and
experimentally tractable, optimum productivity is likely achieved by a
dynamic strategy, in which the partition of resources between biomass
and product formation varies with time \citep{Venayak:2015if}.

Experimental studies have investigated the use of multi-stage
fermentations to increase productivity and/or yield. These techniques
range from simple manipulations, such as changing from aerobic to
anaerobic conditions \citep{Andersson:2007kx}, to more complex genetic
toggle switches \citep{Soma:2014db} or otherwise inducible gene
expression \citep{ValdezCruz:2010ht}. Computational methods have also
been developed to aid in optimizing two-stage fermentation systems.
Gadkar \emph{et al.} optimized the flux profile of glycerol kinase to
maximize the productivity of glycerol from glucose
\citep{Gadkar:2004hb}. Similarly, Anesiadis and coworkers proposed an
extension to Gadkar's work, where parameters for a quorum-sensing toggle
switch are tuned to provide the optimal repression for a target gene to
maximize ethanol productivity \citep{Anesiadis:2008ia}. However, since
these systems prioritize experimental tractability (by choosing a single
reaction target as their control strategy), they do not guarantee that
the optimized productivity reaches the global maximum.

The calculation of maximum theoretical yield is well established in the
metabolic engineering literature, and its use helps guide strain design
and pathway selection for static knockout and/or overexpression efforts.
Maximum yields are useful even though they are not physically
realizable: they reveal the cofactor balancing and pathway split ratios
necessary to achieve optimum carbon conservation. In an analogous
fashion, an estimate of the maximum theoretical productivity of a batch
culture system would prove equally useful for the growing field of
designing dynamic metabolic interventions. However, due to the
computational burden associated with such an estimate, no generalized
framework for optimizing over all feasible flux profiles currently
exists.

In this article, we present an efficient method for calculating the
maximum theoretical productivity of a batch system based on dynamic
optimization \citep{Biegler:2007kr}. In this approach, a dynamic system
representing the changes in metabolite concentration over time is
represented by a collection of interpolating polynomials that divide the
time domain into a set of finite elements. The polynomials are
constrained to be continuous between each finite element, and at each of
a set of defined points, constraints are imposed to ensure the
derivatives of the interpolating polynomials are consistent with maximum
substrate uptake rates and maintenance requirements. In this manner,
optimal metabolite profiles are found in a single optimization using a
large-scale nonlinear programming solver. Dynamic optimization has
previously been used in metabolic modeling, including in an early study
on dynamic flux balance analysis (DFBA) \citep{Mahadevan:2002en} and in
calculating optimal control for fed-batch fermentations
\citep{Hjersted:2008bf}. However, its widespread usage has been limited
by the number of variables that can be simultaneously considered. In
this study, we remove the explicit mass balance constraints by
calculating elementary flux modes \citep{Schuster:1999wg}, and reduce
the dimensionality of the optimization problem using yield analysis
\citep{Song:2009fn}. Our method accounts explicitly for an arbitrary
number of fermentation stages, and allows metabolic fluxes to change
continuously over the course of the entire simulation. Since
productivity and yield cannot be simultaneously maximized, we
demonstrate how this method can be easily extended to calculate the
complete productivity vs.~yield Pareto frontier. We apply our method to
succinic acid production in two microbial hosts: engineered
\emph{Escherichia coli} and native \emph{Actinobacillus succinogenes}.
Succinic acid is a precursor to commodity chemicals in several
industries, and a promising intermediate in the development of
sustainable chemical production routes \citep{Song:2006dy}. We show that
the method can be useful in strain choice by comparing optimal
productivity-yield surfaces for two organisms, and further demonstrate
that nearly optimal yields and productivities can be achieved with only
two discrete flux stages.

\section{Methods}\label{methods}

\subsection{Dynamic flux balance
analysis}\label{dynamic-flux-balance-analysis}

In flux balance analysis (FBA) models, intracellular metabolic reactions
are represented by a stoichiometric matrix \(\mathbf{S}\), such that
\(S_{ij}\) represents the quantity of metabolite \(j\) produced (or
consumed) by reaction \(i\). Fluxes through each reaction are
represented by the vector \(\mathbf{v}\). It is assumed that the time
scales associated with intracellular metabolite equilibria are much
faster than those associated with cell growth or changes to external
metabolite concentrations, and therefore that metabolites in the cell
can be modeled using a pseudo-steady-state approximation,
\(\mathbf{Sv} \approx \mathbf{0}\).

FBA models can be extended to consider the dynamics of substrate
consumption and biomass formation by including specific bounds on
exchange fluxes and allowing the accumulation of depletion of external
metabolites. The dynamic system for cell growth considered in this paper
is
\begin{equation} \frac{dx_i(t)}{dt} = v_i (t) \; x_0(t) \quad \mathrm{for}\; i \in [0, N_X], \label{eq:ode}\end{equation}
where \(\mathbf{x}(t)\) represents the concentration of external
boundary species, chosen such that \(x_0(t)\) is the concentration of
dry cell weight (DCW, in grams), and \(\mathbf{v}(t)\) represents the
flux through each reaction in the cell (in
\(\mathrm{mmol}\; \mathrm{g}_{\mathrm{DCW}}^{-1}\; \mathrm{hr}^{-1}\)),
chosen such that \(v_0(t)\) through \(v_N\) represent the exchange flux
for species \(x_0\) through \(x_N\), respectively. The flux through the
first reaction is therefore the specific growth rate
(\(v_0(t) \equiv \mu(t)\)), and is in units of hr\textsuperscript{-1}.
The resulting dynamic flux balance analysis (DFBA) model takes the form
of an ordinary differential equation with an embedded linear program,
for which efficient methods for the solution of the initial value
problem have been developed \citep{Hoffner:2012hu}.

The objective considered in this paper is to maximize productivity of
the desired metabolite, \(x_p\), by finding the optimum intracellular
fluxes profiles \(\mathbf{v}(t)\) and final fermentation time \(t_f\);
subject to steady-state constraints, reaction reversibility, and
substrate uptake: \begin{equation}\begin{aligned}
\max_{\mathbf{v}(t), t_f} &\frac{x_p(t_f) - x_p(t_0)}{t_f} \\
\mathrm{such \ that}\quad &\mathbf{S} \mathbf{v}(t) = 0, \\
&\mathbf{v_{lb}}(t) \le \mathbf{v}(t) \le \mathbf{v_{ub}}(t)\\
\end{aligned}\label{eq:opt1}\end{equation} The optimization in
eqns.~\ref{eq:ode}, \ref{eq:opt1} therefore takes the form of an optimal
control problem, which can be solved by discretizing in time and solving
the problem using dynamic optimization. In previous applications of
dynamic optimization in DFBA models, intercellular fluxes are either
optimized directly \citep{Hjersted:2008bf} or replaced with
representative input-output reactions found via pathway analysis
\citep{Mahadevan:2002en}. In this study, we take advantage of the
productivity objective to select Pareto-optimal pathways, greatly
reducing the dimensionality of the optimization.

\subsection{Calculation of elementary flux
modes}\label{calculation-of-elementary-flux-modes}

To select the optimal metabolic pathways, we calculate elementary flux
modes (EFMs) for both of the considered metabolic networks. An EFM is a
vector \(\mathbf{r}\) in the right nullspace of \(\mathbf{S}\)
(\(\mathbf{S}\mathbf{r} \equiv 0\)), such that no other elementary mode
has nonzero entries that are a subset of the nonzero entries of \(r_i\)
\citep{Schuster:1999wg, Jevremovic2010}. EFMs contain the important
property that any vector in the right nullspace of \(\mathbf{S}\),
i.e.~any feasible steady-state flux, can be expressed as a nonnegative
combination of the elementary modes \(\mathbf{r}\) \citep{Terzer2008}.
Because of this property and because we are only interested in the
effect of an elementary mode on the maximum productivity, we can
condense the complete set of EFMs to a convex hull in the projection in
which we are interested \citep{Song:2009fn}. We further restrict our
analysis to the Pareto front of EFMs which allow optimum productivity,
removing inefficient modes without affecting the optimal solution.

\subsection{Dynamic optimization}\label{dynamic-optimization}

We find the flux profiles that achieve the maximum productivity via
orthogonal collocation on finite elements, a method for solving
end-point problems involving a dynamic system without an embedded ODE
integrator \citep{Biegler:2007kr}. In the method, the dynamic system is
represented by a series of algebraic constraints that implement an
implicit Runge-Kutta method. The resulting sparse nonlinear program is
then solved via a large-scale nonlinear programming (NLP) solver.
Following the approach of orthogonal collocation \citep{Biegler:2010gt},
we represent the state variables \(\mathbf{x}(t)\) from eq.~\ref{eq:ode}
as a collection of Lagrange interpolating polynomials. A summary of the
dimensions and variables used in this method are presented in
Tables~\ref{tbl:dimen}, \ref{tbl:var}.

\begin{table}[ht]
\centering

\caption{\label{tbl:dimen}Dimensions of the NLP problem. }

\begin{tabular}{@{}lll@{}}
\toprule

Dimension & Index & Description \\\midrule

\(N_X\) & \(i\) & Number of state variables \\
\(N_K\) & \(j\) & Number of finite elements \\
\(N_D\) & \(k\) & Degree of collocation polynomials \\
\(N_F\) & \(l\) & Number of fermentation stages \\
\(N_R\) & \(m\) & Number of elementary flux modes \\

\bottomrule
\end{tabular}

\end{table}

\begin{table}[ht]
\centering

\caption{\label{tbl:var}Variable matrices optimized by nonlinear
programming. These matrices are flattened to a single parameter vector
prior to being passed to IPOPT. }

\begin{tabular}{@{}lll@{}}
\toprule

Variable & Shape & Description \\\midrule

\(\mathbf{X}\) & \((N_X, N_K, N_D)\) & Nodes of the Legendre polynomials
interpolating \(\mathbf{x}(t)\) \\
\(\mathbf{Y}\) & \((N_F, N_R)\) & Fractional expression of each
elementary mode by stage \\
\(\mathbf{A}\) & \((N_K, N_D)\) & Total flux at each collocation
point \\
\(\mathbf{h}\) & \((N_F)\) & Length of each of the finite elements
within a stage \\

\bottomrule
\end{tabular}

\end{table}

The time domain \(t\) of the fermentation is divided into \(N_K\) finite
elements with a non-dimensionalized internal time \(\tau \in [0, 1]\).
Within each finite element, we represent the state vector by a
polynomial of degree \(N_D\) at \(N_D + 1\) collocation points, denoted
\(\tau_n\). We use Gauss-Radau collocation for its stiff decay, and
therefore for \(N_D = 5\), we set
\(\tau_k \in \left\{0, 0.06, 0.28, 0.58, 0.86, 1.00\right\}\):
\begin{equation}
\begin{aligned}
\mathbf{x}_j(\tau) &= \sum_{k=0}^{N_D}\boldsymbol{\ell}_k(\tau)\mathbf{x}_{jk}\\
\mathrm{where}\quad\boldsymbol{\ell}_k(\tau) &= \prod_{n=0, \; n \ne k}^{N_D} \frac{\tau - \tau_{n}}{\tau_k - \tau_{n}}
\end{aligned}
\label{eq:lagrange}\end{equation} To model changes in the flux
distribution, the finite elements are allocated between \(N_F\) distinct
fermentation stages. Within each stage we assume the fractional
distribution of elementary modes is held constant while the total flux
through the system is allowed to vary. The dynamic system can thus be
calculated by: \begin{equation}
\frac{d\mathbf{x}_{ij}}{dt} = \frac{1}{h_l}\frac{d\mathbf{x}_{ij}}{d\tau} = a_{jk}  x_{0jk} \mathbf{R} \; \mathbf{y}_l
\label{eq:coll_dyn}\end{equation} in which \(\mathbf{R}\) is a matrix of
shape \((N_R, N_X)\) that contains the chosen elementary modes;
\(\mathbf{y}_l\) contains the fractional distribution of each elementary
mode in the given stage; \(a_{jk}\) is the time-varying combined flux
through all elementary modes; and \(x_{0jk}\) represents the current
biomass concentration at the given collocation point. In addition to
changes in fractional EFM distribution, the step size \(h_l\) is also
optimized by the nonlinear program and allowed to vary between
fermentation modes.

Solution of the problem involves the constrained optimization over the
variables found in Table~\ref{tbl:var} of the productivity objective
\begin{equation}
\max_{\mathbf{X}, \mathbf{Y}, \mathbf{A}, \mathbf{h}} \frac{x_{p,N_K}(1) - x_{p,0}(0)}{\sum_{j=0}^{N_K} \mathbf{h}(j)}.
\label{eq:coll_obj}\end{equation} Orthogonal collocation imposes a
number of constraints during the optimization process
(eqns.~\ref{eq:dynamics_constraint}, \ref{eq:cont}, \ref{eq:bounds}, \ref{eq:stepsize}, \ref{eq:glcconsump}).
The first of these constraints is that state variable profiles must obey
system dynamics, accomplished by ensuring that the derivative of the
interpolating polynomial defined in eq.~\ref{eq:lagrange} is equal to
the dynamics defined in eq.~\ref{eq:coll_dyn} at each collocation point.
Limits are also placed on the specific uptake and secretion rates at
each collocation point in accordance with biological measurements, and
the sum of the fractional expression of each elementary mode for the
given fermentation stage is constrained to unity. \begin{equation}
\begin{aligned}
\mathrm{for} \; &j \in [0, N_K];\; k \in [1,N_D]:\\
&h_l a_{jk} x_{0jk} \mathbf{R} \; \mathbf{y}_l - \sum_{n=0}^{N_D}\frac{d\boldsymbol{\ell}_n(\tau_k)}{d\tau} \mathbf{x}_{jn} = 0\\
&\mathbf{v_{lb}}(\mathbf{x}_{jk}) \le a_{jk} \mathbf{R} \; \mathbf{y}_l \le \mathbf{v_{ub}}(\mathbf{x}_{jk}) \\
&\sum_m^{N_R} y_{lm} = 1
\end{aligned}
\label{eq:dynamics_constraint}\end{equation} Furthermore, the finite
elements are constrained to be continuous: \begin{equation}
\mathbf{x}_j(1) - \mathbf{x}_{j+1}(0) = 0 \quad \mathrm{for} \; j \in [0, N_K-1]
\label{eq:cont}\end{equation} Bounds are also placed on each of the
optimization variables: \begin{equation}
\begin{aligned}
&0 < \mathbf{X} < 1000 \\
&0 < \mathbf{Y} < 1 \\
&0 < \mathbf{A} < \infty \\
&0.1 < \mathbf{h} < 30
\end{aligned}
\label{eq:bounds}\end{equation}

Finally, two additional constraints are imposed in order to avoid
numerical instability and trivial solutions. The relative change in step
size between fermentation modes is constrained to a factor of 10,
\begin{equation}
\frac{\mathbf{h}_{\max}}{\mathbf{h}_{\min}} \le 10,
\label{eq:stepsize}\end{equation} and the fermentation must consume at
least \(80\%\) of the initial glucose: \begin{equation}
\frac{x^f_{glc}}{x^0_{glc}} \le 0.2
\label{eq:glcconsump}\end{equation}

\subsection{Implementation}\label{implementation}

Metabolic models are specified using cobrapy \citep{Ebrahim:2013gj}.
EFMs for each metabolic model are calculated via efmtool
\citep{Terzer2008}. All calculations were performed in Python, using the
CasADi library \citep{Andersson:2013wo} to implement the orthogonal
collocation method. IPOPT \citep{Wachter:2005hk} was used to solve the
resulting NLP problem.

\section{Results and discussion}\label{results-and-discussion}

\begin{figure}[htbp]
\centering
\includegraphics[width=6.50000in]{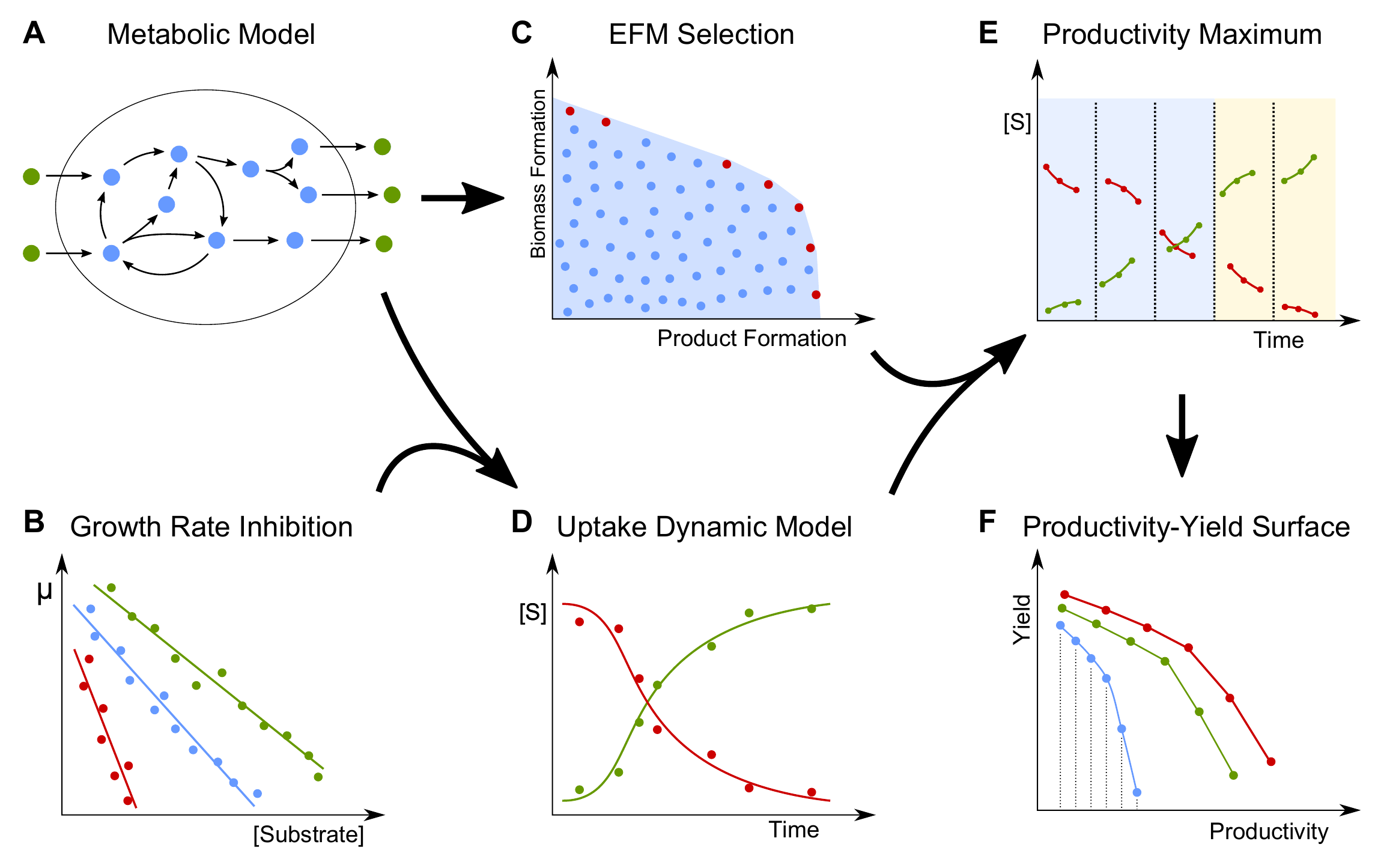}
\caption{\label{fig:comp_summary}\textbf{Method Overview.} A flow chart
for how productivity-yield surfaces are calculated. Stoichiometric
metabolic models (\emph{A}) are generated from literature sources or
genome annotations and are used to calculate the extreme elementary
modes of the system (\emph{C}). Data from growth experiments (\emph{B})
are used together with stoichiometric models to fit dynamic expressions
for substrate uptake kinetics and cellular growth (\emph{D}). Bounds on
substrate uptake are used with the set of EFMs in dynamic optimization
(\emph{E}) in order to calculate optimal flux profiles. This procedure
is repeated to maximize product yield for a number of specified
productivity constraints (\emph{E}), generating the surface of maximum
yield as a function of desired productivity.}\label{fig:compux5fsummary}
\end{figure}

We demonstrate the usefulness of the method by calculating the maximum
theoretical productivities of succinic acid from glucose for two
organisms, \emph{E. coli} and \emph{A. succinogenes}. In a developing a
DFBA model, we integrate knowledge of the core-carbon metabolic pathways
present in the organism together with experimental data on biomass
production and substrate uptake rates. An overview of the computational
method is provided in Figure~\ref{fig:comp_summary}.

\subsection{\texorpdfstring{Core-carbon models of \emph{A. succinogenes}
and \emph{E.
coli}}{Core-carbon models of A. succinogenes and E. coli}}\label{core-carbon-models-of-a.-succinogenes-and-e.-coli}

\subsubsection{\texorpdfstring{\emph{A.
succinogenes}}{A. succinogenes}}\label{a.-succinogenes}

A stoichiometric model of core-carbon metabolism in \emph{A.
succinogenes} was created based on genomic evidence and a previous
metabolic reconstruction \citep{McKinlay2010}. The model consists of 72
metabolites and 89 mass- and charge-balanced reactions. Precursor
metabolites for biomass synthesis were taken from an \emph{E. coli} core
carbon model \citep{Orth:2010eq}. From experimental measurements on the
specific growth and glucose uptake rates of \emph{A. succinogenes},
\(0.414\;\mathrm{hr}^{-1}\) and \(9.5\; \mathrm{mmol}\) respectively
\citep{McKinlay2005}, we tailored the growth-assisted maintenance and
redox cofactors terms in the biomass equation to match known values. No
oxygen uptake was allowed to simulate anaerobic growth. A
non-growth-assisted maintenance value of
\(4.7\;\mathrm{mmol}\; \mathrm{g}_{\mathrm{DCW}}^{-1} \mathrm{hr}^{-1}\)
was also enforced, which was determined from the value estimated by Lin
and coworkers (2008) by finding the amount of ATP that can be produced
from \(0.308\) g of glucose \citep{Lin:2008gm}.

Dynamic constraints on substrate uptake were also imposed. Substrate and
product inhibition on growth rate in \emph{A. succinogenes} have
previously been quantified with a Han-Levenspiel model
\citep{Lin:2008gm}. Using the calculated biomass yield of
\(0.044\; \mathrm{g}_{\mathrm{DCW}}\) / mmol glucose, the experimentally
determined parameters were converted to constraints on glucose uptake:
\begin{equation}
    v_{glc, \min} = \frac{v_{\max}\;x_{glc}}{x_{glc} + K_s}\;\prod_n \left(1 - \frac{x_n}{C_{n}^\star}\right)^{a_n}
\label{eq:glc_uptake}\end{equation} The parameters used in the
eq.~\ref{eq:glc_uptake} are presented in Table~\ref{tbl:lin_parameters}.
Additionally, the lower bound of the flux through non-growth associated
ATP maintenance was constrained to the estimated value: \[
v_{atp, \min} = 4.7
\]

\begin{table}[ht]
\centering

\caption{\label{tbl:lin_parameters}Parameters for the maximum specific
glucose uptake, adapted from values estimated by Lin \emph{et al.}
\citep{Lin:2008gm} }

\begin{tabular}{@{}lll@{}}
\toprule

Parameter & Value & Units \\\midrule

\(v_{\max}\) & \(-11.47\) & \(\mathrm{mmol}\; \mathrm{g}_{\mathrm{DCW}}^{-1}\) \\
\(K_s\) & \(11.27\) & mM \\
\(C^\star_{\mathrm{glucose}}\) & \(860.4\) & mM \\
\(C^\star_{\mathrm{succinate}}\) & \(385.7\) & mM \\
\(C^\star_{\mathrm{formate}}\) & \(235.3\) & mM \\
\(C^\star_{\mathrm{acetate}}\) & \(538.8\) & mM \\
\(a_{\mathrm{glucose}}\) & \(.603\) & - \\
\(a_{\mathrm{succinate}}\) & \(1\) & - \\
\(a_{\mathrm{formate}}\) & \(1\) & - \\
\(a_{\mathrm{acetate}}\) & \(1\) & - \\

\bottomrule
\end{tabular}

\end{table}

\subsubsection{\texorpdfstring{\emph{E. coli}}{E. coli}}\label{e.-coli}

A core-carbon model for \emph{E. coli} central metabolism was taken from
Orth \emph{et al.} \citep{Orth:2010eq}, which consists of 72 metabolites
and 94 reactions. Dynamic models for glucose uptake in \emph{E. coli}
typically assume Michaelis-Menten kinetics, in which high concentrations
of glucose ultimately saturate the import mechanisms at their maximum
value \citep{Hanly:2011il, Song:2010iq, Harwood:2015if}. However, this
assumption leads to the erroneous result that maximum productivity is
achieved at an infinite initial glucose concentration. Since no suitable
literature model for substrate-level growth inhibition from the
considered substrates could be found, glucose uptake kinetics were
adapted from those used for \emph{A. succinogenes}. Additional bounds on
ATP maintenance and oxygen uptake were adapted from those determined by
Feist and coworkers \citep{Feist2007}: \[
\begin{aligned}
v_{atp, \min} &= 8.39\\
v_{o2, \min} &= -18.2\\
\end{aligned}
\]

\subsection{\texorpdfstring{Calculation and selection of elementary
models for \emph{A. succinogenes} and \emph{E.
coli}}{Calculation and selection of elementary models for A. succinogenes and E. coli}}\label{calculation-and-selection-of-elementary-models-for-a.-succinogenes-and-e.-coli}

Elementary flux modes were calculated to reduce the dimensionality of
the optimization and to alleviate the necessity of enforcing
stoichiometric equilibrium constraints during the dynamic optimization.
For \emph{A. succinogenes}, 4763 EFMs were calculated for growth on
glucose and normalized by the~glucose uptake rate. After reducing EFMs
to the considered boundary species, duplicate EFMs were removed. The
boundary species considered for \emph{A. succinogenes} included biomass,
ATP, glucose, succinate, formate, acetate, and pyruvate. Since any
feasible flux state can be expressed as a nonnegative combination of
elementary flux modes, the flux space in the reduced dimensionality can
be spanned without loss of generality by only those EFMs at the vertices
of a convex hull. Furthermore, as optimal succinate productivity will be
achieved using flux modes on the Pareto front of cell growth, ATP
production, and succinate secretion, the reduced set of EFMs were
further reduced to a final set of 22 Pareto-optimal modes. As EFMs are
normalized by glucose consumption, glucose needs not be explicitly
included in the Pareto surface. For \emph{E. coli}, the additional
aerobic growth modes led to a total of \(100,273\) EFMs. The boundary
species for \emph{E. coli} were the same as those used for \emph{A.
succinogenes}, with the addition of oxygen and the omission of pyruvate,
which was not found to be present in any optimal growth mode. In
addition to biomass, ATP, and succinate production, the convex hull of
optimal modes in \emph{E. coli} must also consider oxygen consumption.
After reduction, 137 modes were kept for \emph{E. coli}.

\begin{figure}[htbp]
\centering
\includegraphics[width=6.50000in]{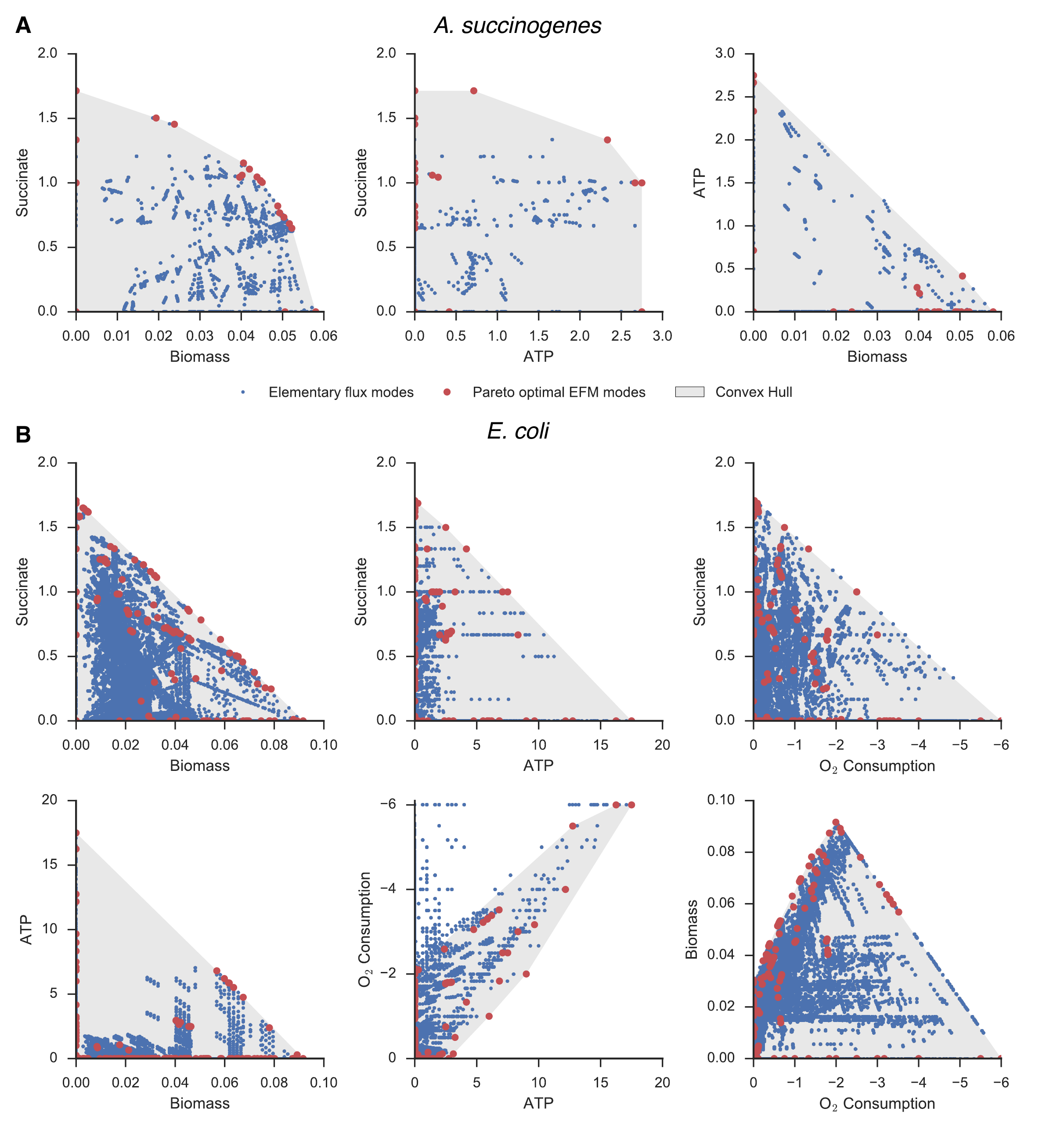}
\caption{\label{fig:EFM_surfs}\textbf{Elementary Flux Mode Surfaces.}
Projections of the yield surfaces are shown for the 3D yield surface of
\emph{A. succinogenes} (\emph{A}) and 4D surface of \emph{E. coli}
(\emph{B}). Elementary flux modes are normalized by glucose uptake, and
therefore points in yield space represent the amount of product (in
mols) which can be produced from 1 mol of glucose. Modes which lie along
the Pareto optimal surface are highlighted in red, and the convex hull
spanned by these points is shaded gray. For \emph{A. succinogenes}, in
which succinate production is coupled to optimal growth, the Pareto
frontier of succinate vs.~biomass yield is sharply curved. Compared with
the corresponding plot in \emph{E. coli}, this shape suggests higher
succinate production can be achieved in \emph{A. succinogenes} without a
linear penalty in growth and ATP yield. The entire flux cone need not be
spanned by the selected EFMs, as demonstrated by the lack of high oxygen
consuming - low ATP producing EFMs selected for the \emph{E. coli}
network.}\label{fig:EFMux5fsurfs}
\end{figure}

The selected EFMs for each organism are shown in
Figure~\ref{fig:EFM_surfs}, which plots 2D slices of the 3D (A) and 4D
(B) yield surfaces. In \emph{A. succinogenes}, the chosen elementary
modes can span the entirety of the yield space for the considered
boundary species. In \emph{E. coli}, modes that consume high amounts of
oxygen while yielding low amounts of ATP are omitted by the algorithm,
as they are unlikely to be utilized in the optimal productivity
solution.

\subsection{Orthogonal collocation}\label{orthogonal-collocation}

\begin{figure}[htbp]
\centering
\includegraphics[width=6.50000in]{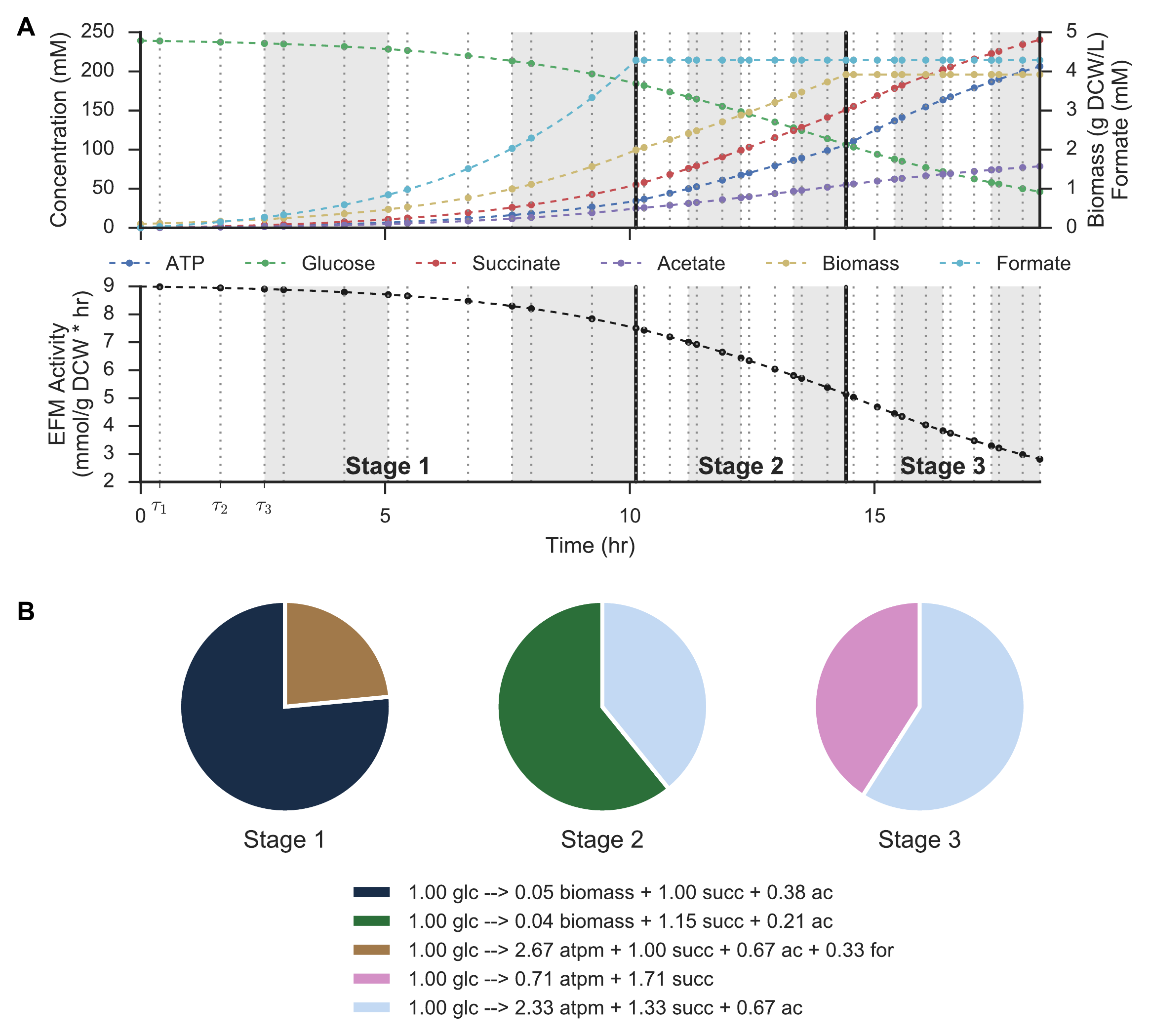}
\caption{\label{fig:coll_schem}\textbf{Schematic of the dynamic
optimization approach.} Example optimum solution for maximum production
of succinate in \emph{A. succinogenes}. Time-course trajectories in
(\emph{A}) are broken down into three stages, each of which is comprised
of a number finite elements (4 in this example, indicated by light/dark
shading). Within each finite elements, state variables are represented
by a lagrange interpolating polynomial at each of 4 collocation points
(\(\tau_0 \rightarrow \tau_3\)), with continuity constraints between
finite elements. The step size of the finite elements, \(h\), within
each stage is also optimized. Metabolic fluxes are optimized by
simultaneously optimizing the fractional expression of each elementary
mode by stage (\emph{B}) with the overall activity of all elementary
modes (\emph{A}, lower plot). In this example, high-growth modes are
replaced by high succinate yielding modes in later fermentation
stages.}\label{fig:collux5fschem}
\end{figure}

Optimum productivities are found via a dynamic optimization framework.
To further reduce the dimensionality of the optimization, and to allow
the effects of discrete fermentation stages to be explicitly simulated,
we divide the fermentation time into a number of discrete stages. Within
each stage, fluxes are represented by a total flux profile through all
elementary modes along with a fractional breakdown of flux through each
elementary mode. An example optimum solution for maximum productivity
from a 3-stage fermentation in \emph{A. succinogenes} is shown in
Figure~\ref{fig:coll_schem}. For clarity, this example uses only 4
finite elements per fermentation stage, while all other calculations use
a minimum of 20 total finite elements evenly divided between stages. The
figure demonstrates how the step sizes \(\mathbf{h}\) are optimized to
achieve the desired dynamic profile. Additionally, since EFMs are scaled
by glucose uptake, the activity parameter smoothly tracks the maximum
allowable uptake rate. ATP production is enforced at each collocation
point by requiring that the flux through the ATP boundary reaction is
greater than the non-growth associated maintenance requirement. The
selected EFMs in Figure~\ref{fig:coll_schem}B demonstrate that optimal
productivity is achieved by prioritizing cell growth early in the
fermentation and succinate production in later stages. The elementary
mode that represents the maximum theoretical yield for succinate on
glucose, 1.71 mol succinate / mol glucose, is used only partially in the
last fermentation stage, highlighting that constraints on cell
maintenance prohibit the system from achieving the maximum theoretical
yield.

\subsection{Effect of increasing the number of fermentation
stages.}\label{effect-of-increasing-the-number-of-fermentation-stages.}

\begin{figure}[htbp]
\centering
\includegraphics[width=6.50000in]{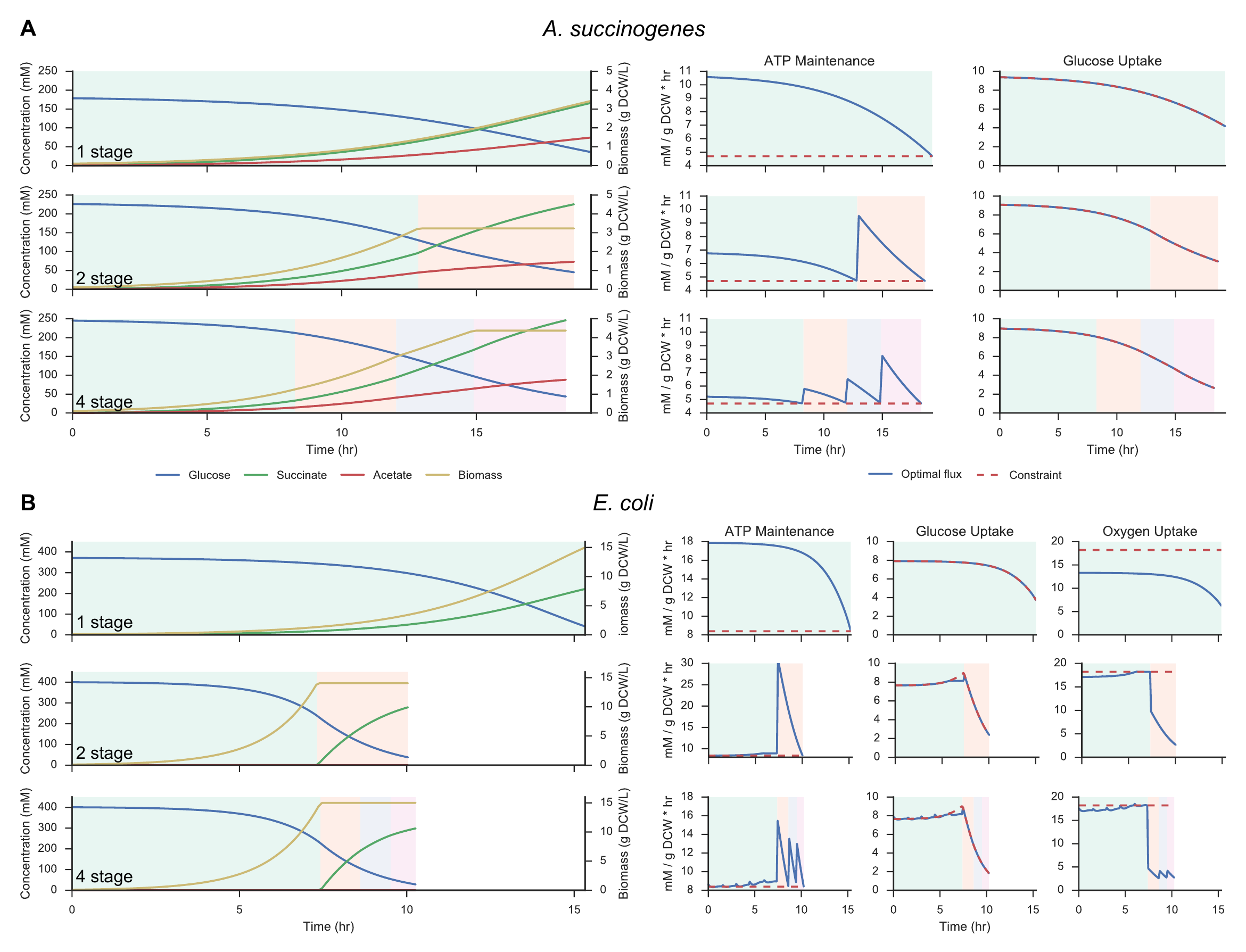}
\caption{\label{fig:multistage_summary}\textbf{Optimal productivity
traces for different numbers of fermentation stages.} (\emph{A}) In
\emph{A. succinogenes}, additional stages achieve higher titers while
only slightly reducing the optimal fermentation time. This result is
achieved by prioritizing growth early in the fermentation and succinate
production during the later stages. Additionally, the greater
flexibility afforded by more stages allows ATP production to be tailored
to meet cellular demand, and less ATP is wasted. (\emph{B}) In \emph{E.
coli}, additional stages drastically reduce the optimum fermentation
time as aerobic growth modes are used to increase biomass ealy in the
fermentation. Additional stages beyond two do not drastically change the
optimum results, indicating that a simple change from aerobic growth to
microaerobic succinate production is close to the global optimum
production strategy.}\label{fig:multistageux5fsummary}
\end{figure}

In addition to reducing the dimensionality of the problem, splitting the
fermentation time into discrete stages with independent EFM expression
allows a systematic investigation into the effect of stage count on
maximum theoretical productivity. Relative flux ratios within each stage
are fixed, and therefore represent consistent enzyme expression. Optimal
productivities achieved for a varying number of fermentation stages are
shown in Figure~\ref{fig:multistage_summary} for \emph{A. succinogenes}
and \emph{E. coli}. The right hand side of
Figure~\ref{fig:multistage_summary} plots the dynamic constraints
imposed on the solution, including ATP maintenance production rate,
maximum glucose uptake rate, and the oxygen uptake rate (for \emph{E.
coli}). In all cases, the solutions closely track the maximum allowable
glucose uptake rate. In \emph{A. succinogenes}, the transition from one
to two stages allows the cell to divide its succinate production
strategy into separate growth and product formation stages. The addition
of subsequent stages permits slightly higher productivities by the
gradual transition from growth to product formation. Additionally, as
the flux ratios are fixed within each fermentation stage, the fraction
of input carbon diverted to ATP production is also fixed. Thus, as lower
extracellular glucose concentrations result in lower glucose uptake
rates, the cell must dynamically allocate a higher percentage of energy
towards meeting its ATP maintenance requirement. Higher numbers of
stages therefore allow the cell to more efficiently meet the maintenance
constraint, as demonstrated by lower excess ATP produced with additional
stages.

In \emph{E. coli}, maximum succinate production using a single stage is
achieved microaerobically. With the addition of a second stage, the
optimal fermentation time drops appreciably as succinate production is
divided into an aerobic growth phase followed by a microaerobic
production phase. The addition of further stages beyond two has little
effect on either the succinate productivity or the concentration
profiles, serving mainly to keep ATP production closer to the
constrained minimum.

\subsection{Pareto surfaces of yield versus
productivity}\label{pareto-surfaces-of-yield-versus-productivity}

While separate estimates of maximum theoretical yield and productivity
can provide information on the economic feasibility of a bioprocess, in
optimized settings one often cannot be increased without decreasing the
other. We therefore show how the given method can be easily extended to
calculate a full productivity-yield Pareto surface: the envelope in the
multi-objective optimization on which productivity cannot be increased
without sacrificing yield. The surface is found by first calculating the
maximum productivity (\(\mathcal{P}_{\max}\), defined by
eq.~\ref{eq:coll_obj}) via the method described previously.

The nonlinear program is then resolved with yield as the objective, \[
\max_{\mathbf{X}, \mathbf{Y}, \mathbf{A}, \mathbf{h}} \frac{x_{p,N_K}(1) - x_{p,0}(0)}{x_{g,0}(0) - x_{g,N_K}(1)},
\] in which \(\mathbf{x}_g\) represents the glucose concentration, while
holding productivity constrained to a fraction of the maximum
productivity, \[
\frac{x_p(t_f) - x_p(t_0)}{t_f} - \alpha\mathcal{P}_{\max} = 0 \quad \mathrm{for}\;\alpha\in[0,1].
\] Computational efficiency in the repeated optimizations is improved by
using IPOPT's warm solve method, which preserves the optimal solution as
a starting guess for the next iteration.

\begin{figure}[htbp]
\centering
\includegraphics[width=5.00000in]{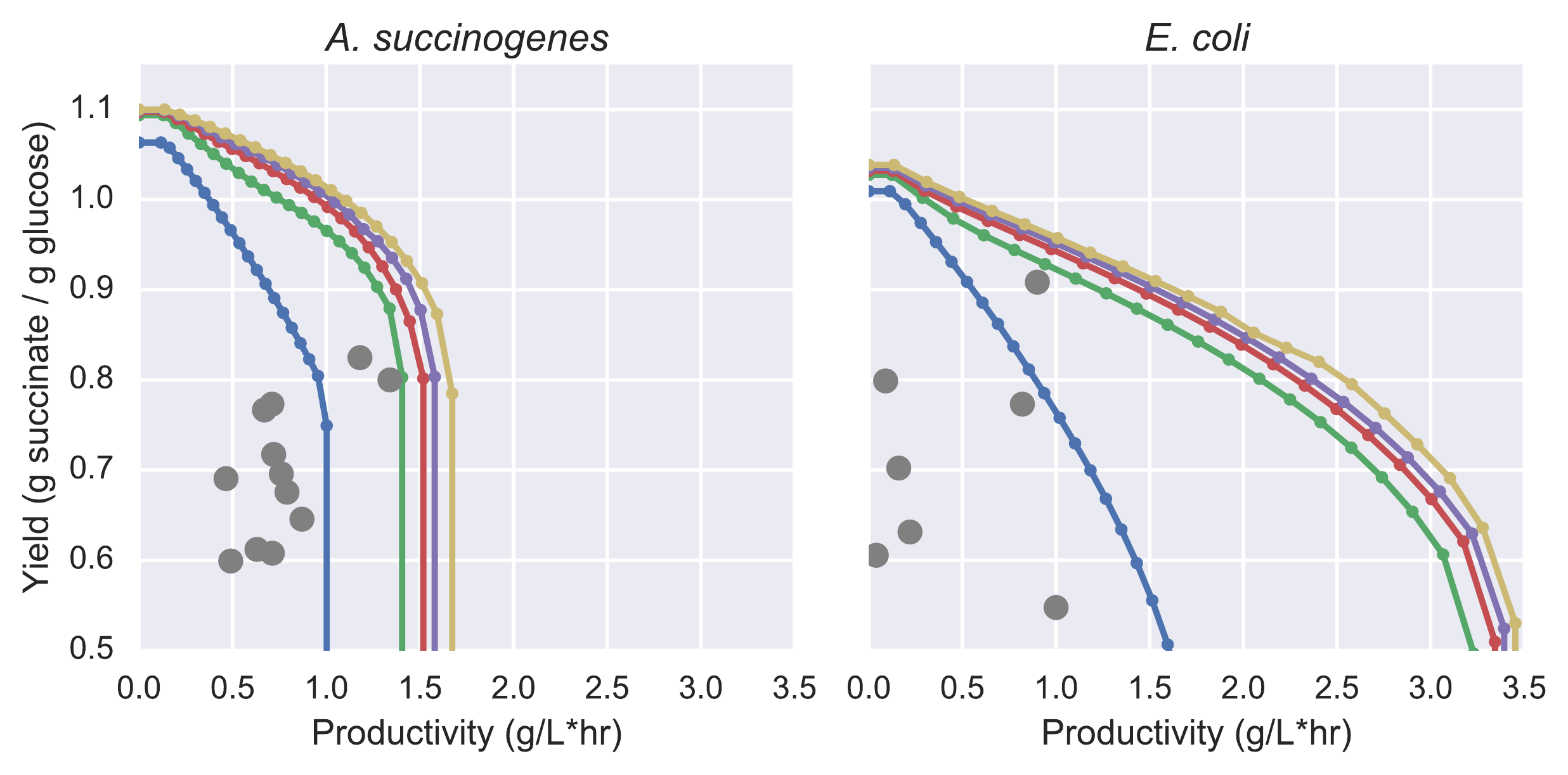}
\caption{\label{fig:yield_surf}\textbf{Productivity-yield pareto
surfaces for \emph{A. succinogenes} and \emph{E. coli}.}
Productivity-yield surfaces are calculated by holding productivity
constant while maximizing yield for a given number of stages.
Experimentally realized productivity and yields are shown in gray, with
most points falling in the space predicted by single-stage strategies.
References for the experimental points are given in
Table~\ref{tbl:supp_succinogenes} and Table~\ref{tbl:supp_ecoli} for
\emph{A. succinogenes} and \emph{E. coli}, respectively. The results
indicate that improving existing yields in \emph{A. succinogenes}
fermentations would lead to greater gains than improving productivity.
Similarly in \emph{E. coli}, the results show that higher productivities
could likely be achieved by leveraging aerobic growth modes more
effectively.}\label{fig:yieldux5fsurf}
\end{figure}

Figure~\ref{fig:yield_surf} plots the resulting yield-productivity
surfaces for 50 linearly spaced \(\alpha\) values. By varying the number
of allowed fermentation stages, these surfaces reveal the performance
gains which can be achieved by allowing additional metabolic
flexibility. For \emph{A. succinogenes}, higher yields can be obtained
due to the lower specific ATP maintenance requirement. Additionally, as
succinic acid is the naturally predominant fermentation product in
\emph{A. succinogenes}, high yields are obtained even at close to the
maximum productivity. In \emph{E. coli}, notable performance gains are
achieved by moving to a 2-stage fermentation, as it enables efficient
usage of aerobic growth modes. In both cases, moving beyond two distinct
flux modes does not substantially increase the yields or productivities
that can be achieved.

The calculated optimal surfaces are compared to data on succinate yield
and productivity that have been achieved experimentally from both
wild-type and engineered organisms, compiled in
Tables~\ref{tbl:supp_succinogenes}, \ref{tbl:supp_ecoli}. The data
largely fall within the predicted envelope of single-stage processes,
confirming the accuracy of the assumptions made in each model. These
results illustrate how the proposed method can be useful for guiding
experimental effort: as wild-type \emph{A succinogenes} fermentations
naturally fall close to the optimal productivity-yield surface, further
genetic manipulation of this organism is unlikely to yield significantly
improved performance. Similarly with \emph{E. coli}, improving succinate
productivity beyond the values previously recorded will likely require a
two-stage process involving aerobic growth and anaerobic succinate
production.

\begin{table}[ht]
\centering

\caption{\label{tbl:supp_succinogenes}Literature data for productivity
and yield achieved in \emph{A. succinogenes} batch fermentations. }

\begin{tabular}{@{}llr@{}}
\toprule

Productivity (g/L*hr) & Yield (g/g) & Reference \\\midrule

1.18 & 0.82 & \citep{Li:2010cd} \\
0.71 & 0.77 & \citep{Li:2010cd} \\
0.49 & 0.60 & \citep{Li:2010cd} \\
0.67 & 0.77 & \citep{Li:2010cd} \\
0.76 & 0.70 & \citep{Li:2010cd} \\
0.63 & 0.61 & \citep{Li:2010cd} \\
0.46 & 0.69 & \citep{Salvachua:2016co} \\
0.72 & 0.72 & \citep{Salvachua:2016co} \\
0.71 & 0.61 & \citep{Salvachua:2016co} \\
0.87 & 0.65 & \citep{Salvachua:2016co} \\
0.55 & 0.76 & \citep{Guettler:1996vz} \\
1.34 & 0.80 & \citep{Guettler:1996vz} \\

\bottomrule
\end{tabular}

\end{table}

\begin{table}[ht]
\centering

\caption{\label{tbl:supp_ecoli}Literature data for productivity and
yield achieved in \emph{E. coli} batch fermentations. }

\begin{tabular}{@{}llr@{}}
\toprule

Productivity (g/L*hr) & Yield (g/g) & Reference \\\midrule

0.90 & 0.91 & \citep{Jantama:2008hx} \\
0.82 & 0.77 & \citep{Jantama:2008hx} \\
1.27 & 0.46 & \citep{Andersson:2007kx} \\
0.60 & 0.28 & \citep{Millard:1996tg} \\
0.22 & 0.63 & \citep{Stols:1997uy} \\
1.00 & 0.55 & \citep{Lin:2005dl} \\
0.16 & 0.70 & \citep{Lin:2005dl} \\
0.09 & 0.80 & \citep{Sanchez:2005cs} \\
0.04 & 0.61 & \citep{Sanchez:2005cs} \\

\bottomrule
\end{tabular}

\end{table}

\section{Conclusions}\label{conclusions-1}

This study represents a computationally efficient method for determining
the maximum theoretical productivity for a batch culture system. As the
fields of metabolic engineering and synthetic biology continue to
develop techniques for the dynamic manipulation of metabolism, our
methodology will enable experimental efforts to be focused on where the
greatest improvements can be expected. While this study has focused on
finding globally optimal solutions, future implementations might search
for optimal strategies using only experimentally tractable EFM
selections (i.e., ones that are growth-optimal for 1 or 2 gene
knockouts). The method can also easily be generalized for conversion of
multiple substrates as long as appropriate experimental data exists to
fit detailed models to substrate uptake kinetics. It could also be
extended to explicitly optimize final titer instead or in addition to
yield and productivity. Overall, this work emphasizes the need for
better empirical models of substrate and product exchange rates and
growth kinetics in designing dynamic metabolic interventions.

\section{Declarations}\label{declarations}

\subsection{Ethics approval and consent to
participate}\label{ethics-approval-and-consent-to-participate}

Not applicable

\subsection{Consent for publication}\label{consent-for-publication}

Not applicable

\subsection{Availability of data and
material}\label{availability-of-data-and-material}

The datasets generated during and/or analysed during the current study
are available in the github repository,
\texttt{https://github.com/pstjohn/productivity-maximization}

\subsection{Competing interests}\label{competing-interests}

The authors declare that they have no competing interests

\subsection{Funding}\label{funding}

This work was funded by the US Department of Energy's Bioenergy
Technologies Office (DOE-BETO), Contract No. DE-AC36-08GO28308 with the
National Renewable Energy Laboratory

\subsection{Authors' contributions}\label{authors-contributions}

PSJ, YJB, and MFC designed the method and wrote the manuscript. PSJ
performed all numerical experiments. All authors read and approved the
final manuscript.

\subsection{Acknowledgements}\label{acknowledgements}

We thank Gregg Beckham for his careful review of the manuscript.

\renewcommand\refname{References}
\bibliography{bibliography.bib}

\end{document}